\documentclass[conference]{IEEEtran}

\usepackage{amsmath}
\usepackage{amssymb}
\usepackage{graphicx}
\usepackage{booktabs}
\usepackage{cite}
\usepackage{hyperref}
\usepackage{verbatim}
\usepackage{array}
\usepackage{multirow}
\usepackage{xcolor}
\usepackage{stfloats}
\usepackage{subcaption}

\title{Optimizing Parameterized Physics-Informed Neural Networks to Solve Multilayered Static Linear Elastic PDEs}

\author{
\IEEEauthorblockN{Hanbo Song$^{*}$}
\IEEEauthorblockN{
\textit{Bellarmine College Preparatory}\\
San Jose, US \\
me@hanbosong.com}
\and
\IEEEauthorblockN{Joseph Lim$^{*}$}
\IEEEauthorblockN{
\textit{Bellarmine College Preparatory}\\
San Jose, US \\
joey@josephdlim.com}
\and
\IEEEauthorblockN{Zhen Zhang}
\IEEEauthorblockA{\textit{Division of Applied Mathematics} \\
\textit{Brown University}\\
Providence, US \\
zhen\_zhang2@brown.edu}
}

\begin{document}

\maketitle

\begin{abstract}
Designing multilayered materials for controlled deformation is heavily bottlenecked by the immense computational costs and time of traditional, industrial-grade finite element methods (FEM). This excessive expense severely limits design space exploration and gradient-based optimization. To streamline the workflow, we propose a framework for parameterized physics-informed neural networks (P2INNs). The framework encompasses three core components: I. a FEM baseline with a Hex8 element and trilinear basis functions, II. a P2INN displacement field model across variable material stiffness and layer thicknesses, and III. FEM-referenced evaluation. The PINN enforces static linear elasticity via Navier-Cauchy residuals with training strategies that prioritize physics, including layerwise PDE decomposition, interface continuity penalty, and compliance aware scaling. For one-layer, pure physics-driven configurations, the PINN achieves a mean volume MAE of 1.56\% and worst-case volume MAE of 2.87\% against the FEM benchmark. For three-layer configurations with controlled supervised training, the PINN achieves a mean volume MAE of 2.53\% and worst-case volume MAE of 4.66\%, meeting the near-5\% worst-case volume-MAE target for preliminary design-space exploration. The trained P2INN model remains a lightweight model with a simple forward pass for future calculations, creating an alternative to traditional FEM. By drastically reducing the need for expensive FEM evaluations, this approach could accelerate the inverse design and optimization of advanced layered architectures for protective structures in various load-heavy or potentially collision-heavy fields.
\end{abstract}


\section{Introduction}
\label{sec:introduction}

The application of physics-informed neural networks (PINNs) to solid mechanics problems has garnered significant attention in coordination with scientific machine learning \cite{raissi2019PINN,karniadakis2021PINNreview}. PINNs, developed by Raissi et al. \cite{raissi2019PINN}, offer the potential to solve partial differential equations (PDEs) without requiring traditional mesh-based discretization, while simultaneously enforcing physical constraints through penalties embedded into the neural network's loss function penalties. However, achieving reliable accuracy for practical engineering problems, particularly those involving parametric variations in geometry and material properties, remains challenging \cite{wang2021understanding,krishnapriyan2021characterizing}. These challenges have been analyzed from multiple perspectives \cite{wang2022expert}.

This paper addresses the specific problem of simulating the behavior of multilayer elastic plate-like structures under surface loading. Such configurations could arise in numerous engineering contexts, including composite materials, protective armor systems, and automobile structural integrity. The core challenge lies in developing a parametric model that can predict mechanical response across variations in Young's modulus and thickness over many layers while maintaining physical consistency and quantitative verification versus successful numerical solutions.

We create a workflow that integrates a trusted and qualified FEM baseline built on linear elasticity theory, a parametrized PINN displacement field model with controlled and designed physics constraints, an automated FEM-referenced evaluation for quantitative assessment, and a lightweight surrogate model for rapid design-space exploration~\cite{liu2022deep}. In this paper, we will discuss the following: A. technical background on FEM and PINNs, B. related works \cite{forrester2008engineering,queipo2005surrogate,gramacy2020surrogates}, their shortcomings, and how work is distinguished from prior solutions, Physics-constrained surrogates without simulation data have been demonstrated \cite{sun2020surrogate}, and physics-constrained learning extends to climate downscaling \cite{zhu2019physics}. C. presenting the problem formulation, D. the specifics of the workflow and detailing PINN architecture with its physics constraints, E. the FEM baseline and evaluation methodology, F. final tuned configuration and results, and end with G. significance, limitations, conclusions, and future works.

\subsection{Finite Element Method for Linear Elasticity}

The finite element method provides the numerical foundation for our verification baseline. For linear elasticity problems, 8-node hexahedral elements with trilinear shape functions represent the standard discretization approach \cite{hughes2012finite,zienkiewicz2013finite}. The strong form of static equilibrium is
\begin{equation}
-\nabla \cdot \boldsymbol{\sigma} = \mathbf{f} \quad \text{in } \Omega,
\end{equation}
where $\mathbf{f}$ is the body force. The weak form of the equilibrium equations is obtained by multiplying the strong form by test functions and integrating over the domain, yielding:
\begin{equation}
\int_{\Omega} \boldsymbol{\sigma} : \boldsymbol{\varepsilon}(\mathbf{v}) \, d\Omega = \int_{\Omega} \mathbf{f} \cdot \mathbf{v} \, d\Omega + \int_{\Gamma_t} \mathbf{t} \cdot \mathbf{v} \, d\Gamma
\end{equation}
where $\mathbf{v}$ represents test functions, $\mathbf{f}$ is the body force, and $\mathbf{t}$ is the traction on the Neumann boundary $\Gamma_t$. Element stiffness matrices are assembled via Gaussian quadrature, with material properties assigned based on element centroid positions for multilayer configurations.

The FEM implementation in our workflow follows the established practices. The element stiffness matrix is computed via Gaussian quadrature with the analytical isoparametric formulation. For multilayer problems, material properties are assigned to elements based on their centroid location relative to layer interfaces. Hughes \cite{hughes2012finite} provides comprehensive overview of linear static and dynamic finite element methodology, while Zienkiewicz et al. \cite{zienkiewicz2013finite} establish the theoretical foundations including convergence analysis and error estimation. The FEM solution, while subject to minor discretization error, provides a trusted baseline against which PINN predictions can be accurately and quantitatively assessed.

\subsection{Physics-Informed Neural Networks}

Physics-informed neural networks \cite{raissi2019PINN}, embed PDE residuals into the loss functions of neural networks during automatic differentiation. The fundamental approach minimizes a composite loss function:
\begin{multline}
\mathcal{L}(\theta) = \int_{\Omega} \bigl| D[u_{\theta}](x) - f(x) \bigr|^2 \, dx \\
+ \lambda \int_{\partial\Omega} \bigl| B[u_{\theta}](x) - g(x) \bigr|^2 \, d\sigma(x)
\end{multline}
where $D$ is the differential operator, $B$ is the boundary operator, and $u_{\theta}$ represents the neural network approximation with parameters $\theta$. In other words, training loss is defined as the sum of the physics loss and boundary loss. This setup enables a mesh-free solution for PDEs while encoding physical constraints directly into the learning process \cite{karniadakis2021PINNreview,lagaris1998ann,weinan2018deep}. Alternative physics-informed architectures include extreme learning machines \cite{dwivedi2020physics}.

The majority of PINN research has focused on fluid mechanics applications \cite{wandel2020learning}. PINNs for hidden fluid mechanics \cite{raissi2020hidden} and NSFnets \cite{jin2021nsfnets} provided special architectures, and they have been thoroughly reviewed \cite{cai2021physics}. Data-driven discovery of governing equations complements direct formulations \cite{luo2020data}, and spectral analysis of turbulence has informed related approaches \cite{stevens2020fourier}.

For solid mechanics applications, PINNs enforce equilibrium equations (such as those present in continuum mechanics, fluid dynamics, and thermodynamics), constitutive relations, and boundary conditions through soft penalty terms. Early work by Lagaris et al. \cite{lagaris1998ann} and Yu et al. \cite{yu2018deep} demonstrated neural network solutions for differential equations, while the Deep Ritz method \cite{weinan2018deep} provided a variational framework for boundary value problems. Energy-based problems \cite{guo2022energy}, hyperelasticity applications \cite{klein2022hyperelastic}, and transfer learning approaches \cite{fang2019deep} have further extended the methodology to nonlinear realms, although the main focus of this paper remains time-independent linear elastic equilibrium. Extensions to stiff problems have been explored \cite{goswami2021learning}, and coupled PDE-PDE systems have also been addressed \cite{bar2022learning}. Over time, PINNs have garnered substantial academic progress regarding their accuracy, including introducing variation energy into the loss terms of the neural network \cite{goswami2020transfer}. Fractional derivatives have been incorporated into physics-informed frameworks \cite{pang2019fPINNs}. But significant challenges remain. Wang et al. \cite{wang2021understanding} and Krishnapriyan et al. \cite{krishnapriyan2021characterizing} demonstrates that PINNs fail in challenging physical setups due to optimization difficulties. These findings motivated our approach with carefully designed loss weights and architecture selection.

\begin{figure*}[!t]
    \centering
    \includegraphics[width=\textwidth]{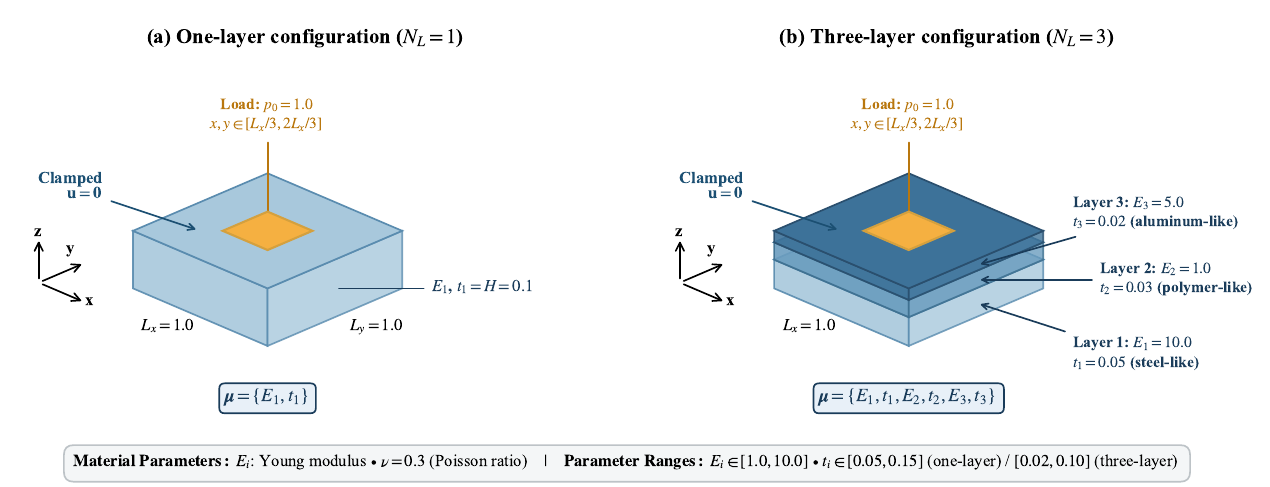}
    \caption{Example problem schematic for a physics-driven parametrized one-layer and three layer elastic plate. (a) One-layer configuration ($N_L = 1$): a single homogeneous plate of thickness $t_1 = H = 0.1$ with Young's modulus $E_1$. (b) Three-layer configuration ($N_L = 3$): a plate partitioned into three layers with varied thicknesses $t_1 = 0.05$, $t_2 = 0.03$, $t_3 = 0.02$ and independent moduli (to resemble real-life materials) $E_1 = 10.0$ (steel-like), $E_2 = 1.0$ (polymer-like), $E_3 = 5.0$ (aluminum-like). Both configurations share the same in-plane dimensions $L_x = L_y = 1.0$, a distributed load $p_0 = 1.0$ over the central top-surface patch, and clamped side boundaries ($\mathbf{u} = \mathbf{0}$). The validated ranges of the parameters are: $E_i \in [1.0, 10.0]$, $t_i \in [0.05, 0.15]$ (one-layer) / $[0.02, 0.10]$ (three-layer), $\nu = 0.3$.}
    \label{fig:problem_setup}
\end{figure*}

\subsection{Parametric PDEs, Neural Operators, and Domain Decomposition}

Parametric PINNs, introduced by Cho et al. \cite{cho2024p2inn}, extend the scope of a traditional PINN by treating material properties, geometric parameters, or boundary conditions as additional network inputs \cite{lu2021deeponet,li2020fourier,cho2024p2inn}. Parametric learning with sparse data has been investigated \cite{howard2023learning}. This enables a single trained model to generalize across a broader, similar family of design spaces, effectively learning to reach an accurate solution over a parameterized domain. However, generalizing over parameters introduces extra complexity since parameters induce significant variations in solution magnitude or spatial structure \cite{wang2021learning}.

Neural operator methods offer an alternative paradigm for parametric PDEs. DeepONet \cite{lu2021deeponet} leverages the universal approximation theorem for operators, using branch and trunk networks to map input functions to output solutions, most with non-linear solutions. Fourier Neural Operators (FNO) \cite{li2020fourier} operate in the spectral domain, achieving computational efficiency through Fast Fourier Transform with $\mathcal{O}(N\log N)$ complexity. Physics-informed DeepONets \cite{wang2021learning} combine operator learning with physics constraints. While these methods excel at learning relations between function spaces and mapping them, they typically require substantial training data and may lack a grounded, physics-based constraint.

For more complex geometries, domain decomposition methods partition the domain into subregions with separate neural networks. Parallel PINNs \cite{shukla2021parallel} assigned different networks to different subdomains and exchange information during training. Combining machine learning with domain decomposition \cite{heinlein2021combining} leverages classical Schwarz and iterative substructuring methods.

\subsection{Boundary and Interface Conditions}

Boundary conditions, in the context of PINNs, are constraints on their solutions spatially and temporally on the boundaries of the domain, which is essential for defining a physical problem. There are two types of BCs enforced: soft constraints \cite{berg2018unified} through the loss function, where penalty terms discourage solutions that violate the BCs, or hard constraints \cite{sukumar2022exact} through modifying the architecture of the network. Enforcing precise BC remains a challenge for PINNs, especially in multilayered structures, where interface conditions between layers present particular challenges. A solution was to strengthen interface sampling with dedicated loss terms, and boundary logic is further improved with PDE decomposition through soft and hard constraints.

\subsection{Error Estimation}

Recent work has addressed error estimation and adaptivity for PINNs. Error estimation frameworks \cite{de2022error} provide posteriori bounds on the solution error. Adaptive sampling strategies \cite{de2022error} refine collocation points in high-error regions. Generalization error estimates \cite{mishra2022estimates} characterize how PINN solutions converge to true PDE solutions as network capacity increases. Our FEM-referenced evaluation protocol provides quantitative error assessment against a baseline. The normalized MAE metric enables comparison across parameter regimes with widely varying displacement magnitudes, providing a scale-invariant evaluation criterion.

\subsection{Training Algorithms and Optimization}

PINN training typically proceeds in two phases: Adam optimization \cite{kingma2014adam} for initial exploration followed by L-BFGS refinement \cite{liu1989limited} for convergence. Adam, the original and universal optimization method, adapts learning rates per parameter using first and second moment estimates, while L-BFGS approximates the Hessian matrix for faster convergence near minima. Self-adaptive strategies \cite{mcclenny2020self} dynamically adjust loss weights during training to balance competing objectives.

Some alternatives include SOAP, developed by Vyas et al. \cite{vyas2024soap}, where the highly effectiveness of Shampoo is coupled with the cheap computational costs of Adam, creating an optimizer that continuously updates the running average of the second moment, but only in the changing coordinate basis. Our work includes the incorporation of the SOAP optimizer for its efficiency and ability to navigate larger model sizes when compared to Adam and its variants.

\section{Problem Formulation}
\label{sec:problem}

\subsection{Domain and Geometry}

We consider a 3D rectangular plate with nominal dimensions
\begin{equation}
L_x = 1.0, \quad L_y = 1.0, \quad H = 0.1
\end{equation}
in normalized units. The domain is partitioned into $N_L$ layers in the thickness direction.

For a one-layer configuration: The total thickness is assigned as a single layer with thickness $t_1$ such that
\begin{equation}
t_1 = H_{\text{total}}.
\end{equation}

Similarly, with a three-layer configuration: The total thickness is partitioned into three layers with thicknesses $t_1$, $t_2$, and $t_3$ such that
\begin{equation}
t_1 + t_2 + t_3 = H_{\text{total}}.
\end{equation}

For both problem setups, each layer is assigned an independent Young's modulus $E_i$, with Poisson's ratio $\nu = 0.3$ held constant across all layers. A distributed load of magnitude $p_0 = 1.0$ is applied over a rectangular patch on the top surface:
\begin{equation}
x \in \left[\frac{L_x}{3}, \frac{2L_x}{3}\right], \quad y \in \left[\frac{L_y}{3}, \frac{2L_y}{3}\right].
\end{equation}

\subsection{Governing Equations}

The governing physics is static linear elasticity. The strain-displacement relation is
\begin{equation}
\boldsymbol{\varepsilon}(\mathbf{u}) = \frac{1}{2}\left(\nabla \mathbf{u} + \nabla \mathbf{u}^T\right),
\end{equation}
where $\mathbf{u} = (u_x, u_y, u_z)$ is the displacement field. The constitutive relation follows Hooke's law:
\begin{equation}
\boldsymbol{\sigma} = \lambda \, \text{tr}(\boldsymbol{\varepsilon}) \mathbf{I} + 2\mu \boldsymbol{\varepsilon},
\end{equation}
with Lam\'{e} parameters
\begin{equation}
\lambda = \frac{E\nu}{(1+\nu)(1-2\nu)}, \quad \mu = \frac{E}{2(1+\nu)}.
\end{equation}
where $\mathbf{\sigma}$ is the stress tensor; $\mathbf{\lambda}$ is Lam\'{e}'s First Parameter, representing the relationship between stress and volumetric strain; and $\mathbf{\mu}$ is Lam\'{e}'s Second Parameter, representing the shear modulus or rigidity modulus of a material.

Static equilibrium requires
\begin{equation}
-\nabla \cdot \boldsymbol{\sigma} = \mathbf{0} \quad \text{in } \Omega.
\label{eq:equilibrium}
\end{equation}

Boundary conditions include: (i) clamped (zero displacement) side faces at $x=0$, $x=L_x$, $y=0$, $y=L_y$; (ii) traction-free top and bottom surfaces outside the load patch; (iii) prescribed normal traction $-p_0$ over the load patch.

\subsection{Parameterization}

One-layer configuration: The PINN accepts spatial coordinates $(x, y, z)$ and a parameter vector
\begin{equation}
\boldsymbol{\mu}_{1L} = [E_1, t_1].
\end{equation}

Three-layer configuration: The PINN accepts spatial coordinates $(x, y, z)$ and a parameter vector
\begin{equation}
\boldsymbol{\mu}_{3L} = [E_1, t_1, E_2, t_2, E_3, t_3].
\end{equation}      

The validated scope focuses on static linear elasticity with $N_L \in \{1, 3\}$ layers, with material and geometric parameters in the ranges $E_i \in [1.0, 10.0]$ and $t_i \in [0.02, 0.10]$ for the three-layer model. The range for Young's modulus remains the same for the one-layer model, but the thickness range becomes $t \in [0.05, 0.15]$. While it is possible to parametrize restitution, friction coefficient, and impact velocity of a specific problem setup in the codebase since the infrastructure accepts additional dynamic parameters, these values are held constant and do not contribute to the verified solution for static elasticity. A complete visualization of both problem setups can be found at Figure~\ref{fig:problem_setup}.

\section{Workflow Architecture}
\label{sec:workflow}

The workflow comprises three interconnected components, depicted conceptually as a processing pipeline.

\subsection{FEM Baseline Generation}

The FEM solver uses 8-node hexahedral (brick) elements with trilinear basis functions. Stiffness matrix assembly follows standard procedures with 2-point Gaussian quadrature in each coordinate direction \cite{hughes2012finite}. Clamped boundary conditions on side faces are enforced via penalty methods. The solver outputs nodal coordinates and displacement fields for verification.

For multilayer configurations, the solver assigns element stiffness matrices based on layer membership determined by element centroid positions. This ensures material discontinuities are captured at layer interfaces. The load patch may use a smooth distribution such as
\begin{equation}
m(x, y) = 16 \cdot \xi(1-\xi) \cdot \eta(1-\eta),
\end{equation}
where $\xi$ and $\eta$ are normalized coordinates within the patch. This smooth distribution represents one example of a patch load formulation that avoids discontinuities.

\subsection{PINN Training}

The PINN is trained to minimize a weighted combination of PDE residual, boundary condition, interface continuity, and optional supervised data losses. Training proceeds in two phases: SOAP optimization for initial exploration followed by L-BFGS refinement for convergence in three-layer model and disabled in the one-layer model \cite{vyas2024soap,kiyani2025optimizing}.

\subsection{Automated Evaluation}

Evaluation performs systematic sweeps over parameter ranges, computing volume mean absolute error (VolMAE) between PINN predictions and FEM reference solutions. The metric is defined as
\begin{equation}
\mathrm{VolMAE}(c)=
100\,
\frac{
\frac{1}{3N_c}\sum_{j=1}^{N_c}
\left\|
\mathbf{u}_{\theta}^{(c)}(\mathbf{x}_j)
-\mathbf{u}_{\mathrm{FEM}}^{(c)}(\mathbf{x}_j)
\right\|_1
}{
\max_{1\le j\le N_c,\;q\in\{x,y,z\}}
\left|u_{\mathrm{FEM},q}^{(c)}(\mathbf{x}_j)\right|
}.
\label{eq:MAE_definition}
\end{equation}

The mean includes all three displacement components at every node of the complete three-dimensional evaluation mesh. This normalization enables meaningful comparison across parameter regimes with widely varying displacement magnitudes.

All quantitative PINN-vs-FEM comparisons use the volume mean absolute error (VolMAE) computed over the same set of evaluation points for a fixed parameter configuration. For configuration $c$, with PINN displacement prediction $\mathbf{u}_{\theta}^{(c)}(\mathbf{x}_j)$ and FEM displacement reference $\mathbf{u}_{\mathrm{FEM}}^{(c)}(\mathbf{x}_j)$ at $N_c$ sampled points, we calculate VolMAE with Equation~\eqref{eq:MAE_definition}.

For a parameter sweep containing $M$ configurations, the reported mean volume MAE and worst-case volume MAE are
\begin{equation}
\mathrm{Mean\ VolMAE}
=
\frac{1}{M}\sum_{c=1}^{M}\mathrm{VolMAE}(c),
\end{equation}
\begin{equation}
\mathrm{Worst\ VolMAE}
=
\max_{1\le c\le M}\mathrm{VolMAE}(c).
\end{equation}
Thus, ``Mean VolMAE'' denotes the average normalized error across the evaluated parameter sweep, while ``Worst VolMAE'' denotes the largest normalized error observed among those configurations. The same definitions are used for the verification and ablation tables below.

\section{PINN Architecture}
\label{sec:PINN}

\subsection{Network Architecture}

The baseline PINN uses a multilayer perceptron with 4 hidden layers of 64 neurons each, with hyperbolic tangent activation. The network predicts a 3-component displacement field.

\subsection{Input Normalization}

Spatial coordinates and material parameters are normalized to $[0,1]$ based on their specified ranges. The through-thickness coordinate is normalized as
\begin{equation}
\hat{z} = \frac{z}{H_{\text{eff}}}
\end{equation}
where $H_{\text{eff}} = t_1$ for the one-layer model and $H_{\text{eff}} = t_1 + t_2 + t_3$ for the three-layer model. In the latter cases, the value will vary on different parametric verifications, but will always be normalized to between 0 and 1.

\subsection{Layerwise Material Handling}

For the three-layer model, material properties vary discontinuously across layer interfaces. Input features encode layer boundaries: normalized signed distances to interfaces
\begin{equation}
\tilde{z}_1 = \frac{z - t_1}{H}, \quad \tilde{z}_2 = \frac{z - t_1 - t_2}{H},
\end{equation}
and soft layer indicators $\sigma(\beta \tilde{z}_i)$ with $\beta=20$. These differentiable indicators enable gradient-based training across material discontinuities.

For physics computations, the constitutive law uses layer-specific stiffness selected by spatial position:
\begin{equation}
E_{\text{local}}(z) =
\begin{cases}
E_1, & z \leq t_1 \\
E_2, & t_1 < z \leq t_1 + t_2 \\
E_3, & z > t_1 + t_2
\end{cases}
\end{equation}

\subsection{Compliance Scaling}

The network output $v$ is converted to physical displacement via a 
\begin{equation}
\mathbf{u} = v \times \frac{1}{E_{\text{eff}}^p} \times \left(\frac{H_{\text{ref}}}{H_{\text{eff}}}\right)^{\alpha}
\end{equation}
where $E_{\text{eff}} = E_1$ (one-layer) or $E_{\text{avg}} = (E_1+E_2+E_3)/3$ (three-layer), $H_{\text{ref}} = 0.1$, with $\alpha = 1.234$, $p = 0.973$ for the one-layer model and $\alpha = 3.0$, $p = 0.95$ for the three-layer model.

\subsection{Boundary Condition Handling}

Side boundaries are clamped ($\mathbf{u} = \mathbf{0}$). Hard enforcement applies a multiplicative mask:
\begin{equation}
\mathbf{u}_{\text{masked}} = 16 \cdot x(1-x) \cdot y(1-y) \cdot \mathbf{u}_{\text{raw}}.
\end{equation}

Both models enforce hard boundary conditions through clamped-side displacements enforcement with hard-masks (analytically zeroing displacement at $x=0$, $x=1$, $y=0$, $y=1$), supplemented by a soft penalty loss. In the design space, these enforcements are enforced on the side walls.

Meanwhile, soft enforcement then adds boundary residual penalties to the loss function to enforce displacement. Traction boundary conditions on the top and bottom surfaces are enforced purely as soft penalties.

\subsection{Loss Functions}

The total loss combines six terms:
\begin{align}
\mathcal{L}_{\text{total}} &= w_{\text{PDE}}\mathcal{L}_{\text{PDE}} + w_{\text{BC}}\mathcal{L}_{\text{BC}} + w_{\text{load}}\mathcal{L}_{\text{load}} \nonumber \\
&\quad + w_{\text{energy}}\mathcal{L}_{\text{energy}} + w_{\text{interface}}\mathcal{L}_{\text{interface}} + w_{\text{data}}\mathcal{L}_{\text{data}}.
\end{align}

PDE Residual Loss. Enforces static equilibrium $\nabla \cdot \boldsymbol{\sigma} = \mathbf{0}$ at interior collocation points:
\begin{equation}
\mathcal{L}_{\text{PDE}} = \frac{1}{N_1}\sum_{i \in \Omega_1} r_i^2 + \frac{1}{N_2}\sum_{i \in \Omega_2} r_i^2 + \frac{1}{N_3}\sum_{i \in \Omega_3} r_i^2,
\end{equation}
where $r_i = -\nabla \cdot \boldsymbol{\sigma}(\mathbf{x}_i) \cdot H_{\text{ref}}$ is the scaled residual and $\Omega_k$ denotes collocation points in layer $k$.

Boundary Condition Loss. Enforces zero displacement on clamped side boundaries ($x = 0, L_x$ and $y = 0, L_y$):
\begin{equation}
\mathcal{L}_{\text{BC}} = \frac{1}{N_{\text{sides}}}\sum_{i=1}^{N_{\text{sides}}} \|\mathbf{u}(\mathbf{x}_i)\|^2, \quad \mathbf{x}_i \in \Gamma_{\text{sides}}.
\end{equation}

Load Patch Traction Loss. Enforces prescribed downward traction $p_0$ on the central load patch:
\begin{equation}
\mathcal{L}_{\text{load}} = \frac{1}{N_{\text{load}}}\sum_{i=1}^{N_{\text{load}}} \|\mathbf{T}(\mathbf{x}_i) - \mathbf{t}_{\text{prescribed}}\|^2, \quad \mathbf{x}_i \in \Gamma_{\text{load}},
\end{equation}
where $\mathbf{T} = \boldsymbol{\sigma} \cdot \mathbf{n}$ is the traction vector and $\mathbf{t}_{\text{prescribed}} = [0, 0, -p_0]^\top$ on the load patch. 

Energy Consistency Loss. Enforces balance between internal strain energy and external work:
\begin{equation}
\mathcal{L}_{\text{energy}} = \left| \frac{1}{2}\langle\boldsymbol{\sigma}, \boldsymbol{\varepsilon}\rangle_{\Omega_h} - (-p_0) \langle u_z \rangle_{\Gamma_{\text{load}}} A_{\text{patch}} \right|,
\end{equation}
where $\langle \cdot \rangle$ denotes the mean over collocation points and $A_{\text{patch}}$ is the load patch area.

Interface Continuity Loss. Enforces continuity of the full traction vector across layer interfaces. 
\begin{equation}
\begin{aligned}
\mathcal{L}_{\text{interface}}
=
\sum_{k=1}^{2}
\mathbb{E}_{\mathbf{x}\in\Gamma_k}
\Big[
&\left(\sigma_{xz}^{(k)}-\sigma_{xz}^{(k+1)}\right)^2 \\
+{}&\left(\sigma_{yz}^{(k)}-\sigma_{yz}^{(k+1)}\right)^2 \\
+{}&\left(\sigma_{zz}^{(k)}-\sigma_{zz}^{(k+1)}\right)^2
\Big],
\end{aligned}
\end{equation}
where $\Gamma_k$ denotes the interface between layers $k$ and $k+1$, and the loss is evaluated at interface collocation points. 

Data Supervision Loss. Penalizes deviation from sparse FEM reference data:
\begin{equation}
\mathcal{L}_{\text{data}} = \frac{1}{N_{\text{data}}}\sum_{i=1}^{N_{\text{data}}} \|\mathbf{v}_{\text{pred}}(\mathbf{x}_i) - \mathbf{v}_{\text{target}}(\mathbf{x}_i)\|^2,
\end{equation}
where $\mathbf{v}_{\text{target}} = \mathbf{u}_{\text{FEM}} / \text{compliance\_scale}(E, t)$ maps physical displacement to the network's compliance-scaled space.

\section{FEM Reference Solver and Evaluation Protocol}
\label{sec:fem_eval}

\subsection{FEM Reference Solver}

The FEM baseline serves as the numerical reference for all verification experiments. The solver uses 8-node brick elements with trilinear shape functions, and the element stiffness matrix is assembled via Gaussian quadrature with the analytical isoparametric formulation. For multilayer configurations, elements are assigned material properties according to their centroid location relative to the layer interfaces. This construction enables piecewise-constant elastic properties across the thickness while preserving a standard linear-elastic finite-element formulation. Although, the solver is not flawless and still remains subject to minor discretization errors. However, these errors remain insignificant, proving the baseline to be a qualified verification metric.

\subsection{FEM Baseline Validation}
\begin{figure*}[!t]
    \centering
    \includegraphics[width=\textwidth]{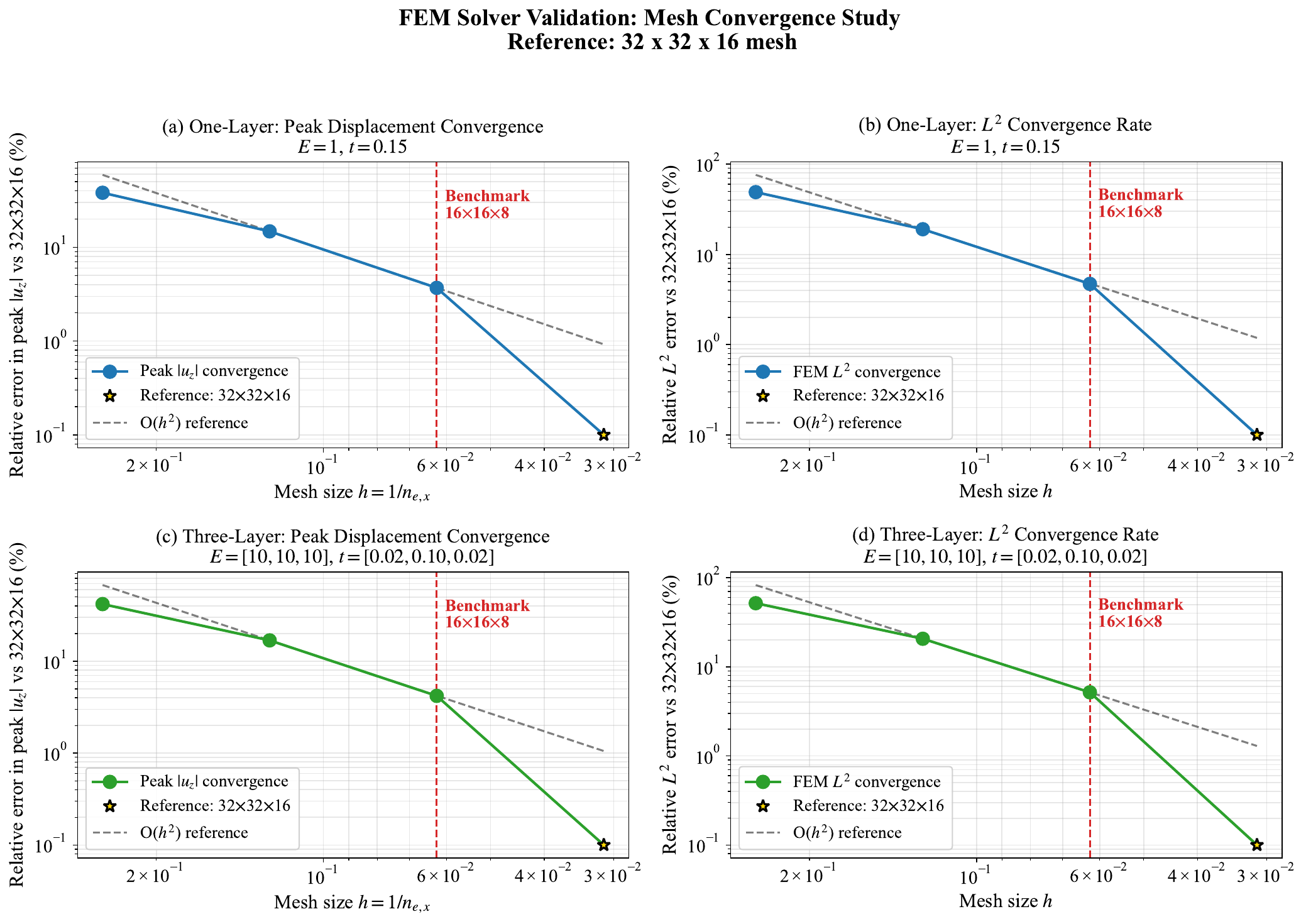}
    \caption{FEM baseline validation: peak displacement and relative $L^2$ error versus mesh size for one-layer and three-layer configurations, demonstrating monotonic convergence.}
    \label{fig:mesh_convergence}
\end{figure*}

To validate the FEM baseline for its accuracy, a mesh convergence study was performed on representative cases. Figure~\ref{fig:mesh_convergence} shows  peak displacement v. mesh refinement, as well as  relative error v. mesh size for both one-layer ($E=10$, $t=0.05$) and three-layer ($E=[10, 10, 10]$, $t=[0.02,0.10,0.02]$) configurations. In both cases, the solution converges monotonically as mesh density increases. The benchmark mesh (16 x 16 x 8) was selected to balance accuracy and computational cost. Subsequent PINN and FEM comparisons are performed on identical meshes to ensure consistency. The study confirms that the solver behaves correctly. Since the convergence error remains low, it can thus be concluded that the FEM model converges successfully and thus can act as a trustworthy baseline.

The relative $L_2$ error metric for the varying FEM mesh sizes are calculated as follows:
\begin{equation}
\|e\|_{L_2}^{\text{rel}} = 100 \times \frac{\sqrt{\int_\Omega (u_h - u_{\text{ref}})^2 \, d\Omega}}{\sqrt{\int_\Omega u_{\text{ref}}^2 \, d\Omega}},
\label{eq:fem_rel_l2}
\end{equation}
where $u_h$ denotes the FEM solution on a mesh with characteristic size $h$, $u_{\text{ref}}$ denotes the FEM solution on the finest reference mesh ($32\times32\times16$), and $d\Omega = dx\,dy\,dz$ is the volume element. On the uniform evaluation grid, this integral is approximated by the discrete sum
\begin{equation}
\int_\Omega f \, d\Omega \approx \sum_{i=1}^{N_x}\sum_{j=1}^{N_y}\sum_{k=1}^{N_z} f(x_i, y_j, z_k) \, \Delta x \Delta y \Delta z.
\label{eq:discrete_sum}
\end{equation}

\subsection{Evaluation Regime}

To assess parametric generalization, we deploy two evaluation protocols. For the one-layer model, 100 random interior configurations were sampled uniformly from
\[
E \in [1.0,10.0], \qquad t \in [0.05,0.15],
\]

For the three-layered model, 100 random interior configurations were sampled from
\[
E_i \in [1.0, 10.0], \qquad t_i \in [0.02, 0.10],
\quad i\in\{1,2,3\}.
\]
Both primary verification studies use 100 uniformly sampled random interior parameter configurations, the fixed random seed 0430, a $16\times16\times8$ element FEM mesh, 2,601 evaluation nodes, and all three displacement components. For each configuration, FEM reference solutions are computed on the same mesh resolution used for representative visualization cases.

Performance is quantified using the volume mean absolute error in Equation~\eqref{eq:MAE_definition}, which enables meaningful comparison across regimes with widely varying displacement magnitudes. Displacement errors are evaluated over all three displacement components at every node of the three-dimensional evaluation mesh.

\subsection{Computational Cost and Efficiency}

A central motivation for the PINN framework is to reduce the computational costs over repeated testing in conventional FEM solvers. Since they require a new calculation for every test case rather than a simple forward pass post-training, FEM's high-level costs lie in its inability to be quickly reused. 

The FEM baseline solves a linear system for each distinct parameter configuration. For the $16 \times 16 \times 8$ hexahedral mesh used in verification (2048 elements, 17x17x9 = 2601 nodes before constraints), a single Python-based FEM solve requires stiffness-matrix assembly via $2 \times 2 \times 2$ Gaussian quadrature followed by a direct linear solve. Each such solve constitutes an independent computation.

The PINN, by contrast, front-loads computation into a one-time training phase that could be generalized over a large parameterized design space. The final model is trained for 400 SOAP epochs with $\sim$43{,}000 collocation points per batch on a 4-layer, 64-neuron MLP. Once trained, inference reduces to a single forward pass that evaluates the displacement field at an arbitrary or user-selected parameter configuration. This requires only matrix-vector products through the network, independent of the PDE discretization.

Table~\ref{tab:cost_one_layer} and Table~\ref{tab:cost_three_layer} contrast the two approaches for a single-case evaluation as well as a finer parameter sweep with $10^6$ configurations representative of a practical design exploration. Each configuration is stochastically chosen through Monte Carlo sampling, and the fine sweep test conservatively models that. In all cases, the PINN avoids repeated assembly-solve cycles, which becomes burdensome for a fine-mesh FEM.

\begin{table}[htbp]
\caption{One-Layer Model (Physics-Driven) v. FEM}
\label{tab:cost_one_layer}
\centering
\resizebox{\columnwidth}{!}{%
\begin{tabular}{@{}lcc@{}}
\toprule
\textbf{Stage} & \textbf{FEM} & \textbf{P2INN} \\
\midrule
Per-configuration eval & 1 assembly + 1 solve & 1 forward pass \\
Training (one-time) & --- & 6.2 min \\
Time per eval & 0.47 s & 0.0025 s \\
Post-training cost & $\mathcal{O}(N)$ solves & $\mathcal{O}(N)$ forward passes \\
\midrule
\textbf{Speedup} & & \textbf{188$\times$} \\
\midrule
Fine sweep ($10^6$ configs) & $10^6$ solves & $10^6$ queries \\
Total time & 132 hrs & 0.8 hrs \\
\bottomrule
\end{tabular}%
}
\end{table}

\begin{table}[htbp]
\caption{Three-Layer Model (Data + Physics) v. FEM}
\label{tab:cost_three_layer}
\centering
\resizebox{\columnwidth}{!}{%
\begin{tabular}{@{}lcc@{}}
\toprule
\textbf{Stage} & \textbf{FEM} & \textbf{P2INN} \\
\midrule
Per-configuration eval & 1 assembly + 1 solve & 1 forward pass \\
Training (one-time) & --- & 1.57 min \\
Time per eval & 0.44 s & 0.0030 s \\
Post-training cost & $\mathcal{O}(N)$ solves & $\mathcal{O}(N)$ forward passes \\
\midrule
\textbf{Speedup} & & \textbf{149$\times$} \\
\midrule
Fine sweep ($10^6$ configs) & $10^6$ solves & $10^6$ queries \\
Total time & 122.4 hrs & 0.9 hrs \\
\bottomrule
\end{tabular}%
}
\end{table}

The key efficiency distinction is not just the speed of a single solve, since the coarse FEM mesh remains small, but rather the scaling with design space cardinality. As shown through the temporal comparisons between the two methods over larger numbers of solves, the PINN severely outperforms the FEM for design space fine sweeps. For parametric optimization, where gradient-based search may require thousands to millions of function evaluations, the PINN's millisecond forward pass efficiency replaces a repeated FEM assembly-solve cycle.

It is also important to note that the training interval for the three-layered model is shorter than that of the one-layered model. But this originates from the existence of data supervision that was lacking in the one-layered model. The three-layered model, guided by partial supervision data, requires less training epochs and thus less temporal requirements for convergence.

\subsection{Sparse FEM Supervision}

In addition to physics-based training via sampled collocation points, lightweight FEM supervision is used to stabilize the most extreme compliance regimes. Each of the three layers has two design parameters, Young's modulus $E_i$ and thickness $t_i$. Both parameters are sampled at their extreme values only: $E_i \in \{1.0, 10.0\}$ and $t_i \in \{0.02, 0.10\}$. Supervision is generated for every combination of these per-layer extremes, giving $(E_1, t_1) \times (E_2, t_2) \times (E_3, t_3) = 4^3 = 64$ distinct FEM configurations spanning the corners of the six-dimensional material–thickness parameter space. The supervision budget of 36{,}000 nodal displacement values is allocated across these 64 configurations, with bias toward thinner stacks through a thickness-weighting power of 3.0. Within each FEM solution, sampled nodes are arranged in the following manner: 60\% from the loaded top surface, 30\% from the layer interfaces, and 10\% from the bulk interior. Supervision data is generated once before training and cached for reuse.

In order to prevent the model from becoming over-reliant on data points and thereby degenerating into a pure data-driven surrogate, the FEM supervision is deliberately sparse in both coverage and sampling. Because supervision is provided exclusively at parameter extremes and never at interior parameter combinations (e.g.\ $E=5$ or $t_i=0.06$), the model must rely on learned physics to simulate across the full parameter space, maintaining a balance between a pure physics-driven network versus a pure FEM-based surrogate that  favors the former. The thickness-weighting bias concentrates supervision in thin-stack regimes, where compliance effects are most severe and physics alone is least reliable. This design improves accuracy in high-compliance regimes without turning the method into a predominantly data-driven surrogate.

\section{Verification Results}
\label{sec:verification_results}

\subsection{Verification Objective}

The primary objective is to obtain a parameterized PINN that predicts single-layer and multilayer elastic response across stiffness and thickness extremes while matching the FEM reference of around a 5\% worst-case MAE. This target is intended for preliminary design-space exploration as a precursor to a more in-depth workflow, where accuracy must remain stable even in mechanically challenging regimes. It is also important to note key distinctions between the pure physics-driven training of the one-layer model against the data-supported training of the three-layer model.

\subsection{Effect of the Proposed Training Design}

The final model performance arises from a combination of physics-first design choices rather than pure data fitting and supervision. The most important ingredients are:
\begin{itemize}
    \item layerwise PDE residual decomposition, which balances residual contributions across stiff and soft layers, preventing stiff layers from dominating the soft layers in a unified loss landscape;
    \item strong interface continuity enforcement, which improves displacement consistency across material boundaries and modeling a realistic transfer of force across a multilayer system;
    \item compliance-aware scaling in thickness and stiffness, which better reflects the mechanics of multilayer plate response;
    \item sparse FEM supervision at parameter extremes, which anchors the solution in high-compliance regimes, adhering to a high-fidelity solution; and
    \item strengthened interface sampling, which concentrates training effort where discontinuous material structure creates the greatest difficulty.
\end{itemize}

Table~\ref{tab:config_three_layer} summarizes the final three-layer configuration used.

\begin{table}[!b]
\caption{Final Three-Layer PINN Configuration}
\label{tab:config_three_layer}
\centering
\footnotesize
\setlength{\tabcolsep}{4pt}
\begin{tabular}{@{}lc@{}}
\toprule
\textbf{Parameter} & \textbf{Value} \\
\midrule
\multicolumn{2}{@{}l@{}}{\textbf{Physics Collocation Points (Unsupervised)}} \\
Interior PDE points & 15,000 \\
Side boundary points & 2,000 \\
Top load patch points & 6,000 \\
Top free surface points & 2,000 \\
Bottom surface points & 2,000 \\
Interface points (total) & 16,000 \\
Interface sample fraction & 0.75 \\
\textbf{Total physics points per batch} & $\sim$43,000 \\
\midrule
\multicolumn{2}{@{}l@{}}{\textbf{FEM Supervision Points}} \\
Supervision points & 36,000 \\
FEM mesh for supervision & $16 \times 16 \times 8$ \\
Supervision thickness power & 3.0 \\
\midrule
\multicolumn{2}{@{}l@{}}{\textbf{Loss Weights}} \\
PDE weight & 10.0 \\
Boundary condition weight & 0.7 \\
Load weight & 5.0 \\
Interface continuity weight & 300.0 \\
Data (supervision) weight & 400.0 \\
\midrule
\multicolumn{2}{@{}l@{}}{\textbf{Physics-First Design}} \\
PDE decompose by layer & Enabled \\
Thickness compliance exponent $\alpha$ & 3.0 \\
Stiffness compliance exponent $p$ & 0.95 \\
Displacement compliance scale & 1.0 \\
\bottomrule
\end{tabular}
\end{table}

Table~\ref{tab:config_one_layer} summarizes the final one-layer configurations used.

\begin{table}[!b]
\caption{Final One-Layer PINN Configuration}
\label{tab:config_one_layer}
\centering
\footnotesize
\setlength{\tabcolsep}{4pt}
\begin{tabular}{@{}lc@{}}
\toprule
\textbf{Parameter} & \textbf{Value} \\
\midrule
\multicolumn{2}{@{}l@{}}{\textbf{Physics Collocation Points (Unsupervised)}} \\
Interior PDE points & 15,000 \\
Side boundary points & 2,000 \\
Top load patch points & 6,000 \\
Top free surface points & 2,000 \\
Bottom surface points & 2,000 \\
Interface points (total) & N/A (single layer) \\
Interface sample fraction & N/A (single layer) \\
\textbf{Total physics points per batch} & $\sim$27,000 \\
\midrule
\multicolumn{2}{@{}l@{}}{\textbf{FEM Supervision Points}} \\
Supervision points & Disabled \\
FEM mesh for supervision & N/A \\
\midrule
\multicolumn{2}{@{}l@{}}{\textbf{Loss Weights}} \\
PDE weight & 10.0 \\
Boundary condition weight & 0.7 \\
Load weight & 5.0 \\
 Interface continuity weight & N/A \\
Data (supervision) weight & 0.0 \\
\midrule
\multicolumn{2}{@{}l@{}}{\textbf{Physics-First Design}} \\
PDE decompose by layer & N/A \\
Thickness compliance exponent $\alpha$ & 1.234 \\
Stiffness compliance exponent $p$ & 0.973 \\
Displacement compliance scale & 1.0 \\
\bottomrule
\end{tabular}
\end{table}

\subsection{PINN-vs-FEM Quantitative Results}

Table~\ref{tab:PINN_results} reports the PINN-vs-FEM verification results for the one-layer and three-layer configurations, measured using the volume-MAE definitions above.

\begin{table}[htbp]
\caption{PINN-vs-FEM verification results.}
\label{tab:PINN_results}
\centering
\begin{tabular}{lcc}
\toprule
\textbf{Configuration} & \textbf{Mean Volume MAE (\%)} & \textbf{Worst Volume MAE (\%)} \\
\midrule
One-layer & 1.56 & 2.87 \\
Three-layer & 2.53 & 4.66 \\
\bottomrule
\end{tabular}
\end{table}

For one-layer configurations, the model achieves a sweep mean volume MAE of 1.56\% and a sweep worst-case volume MAE of 2.87\%. For three-layer configurations, the final model achieves a sweep mean volume MAE of 2.53\% and a sweep worst-case volume MAE of 4.66\%, satisfying the near-5\% worst-case volume-MAE target. The worst three-layer case occurs at
\[
E=[1.201, 9.348, 7.196],
\]
\[
t=[0.06322, 0.02582, 0.05095],
\]

And the worst one-layer case occurs at 
\[
E = 9.203, \qquad t = 0.05375.
\]

\begin{figure}[!t]
\centering

\includegraphics[width=\columnwidth]{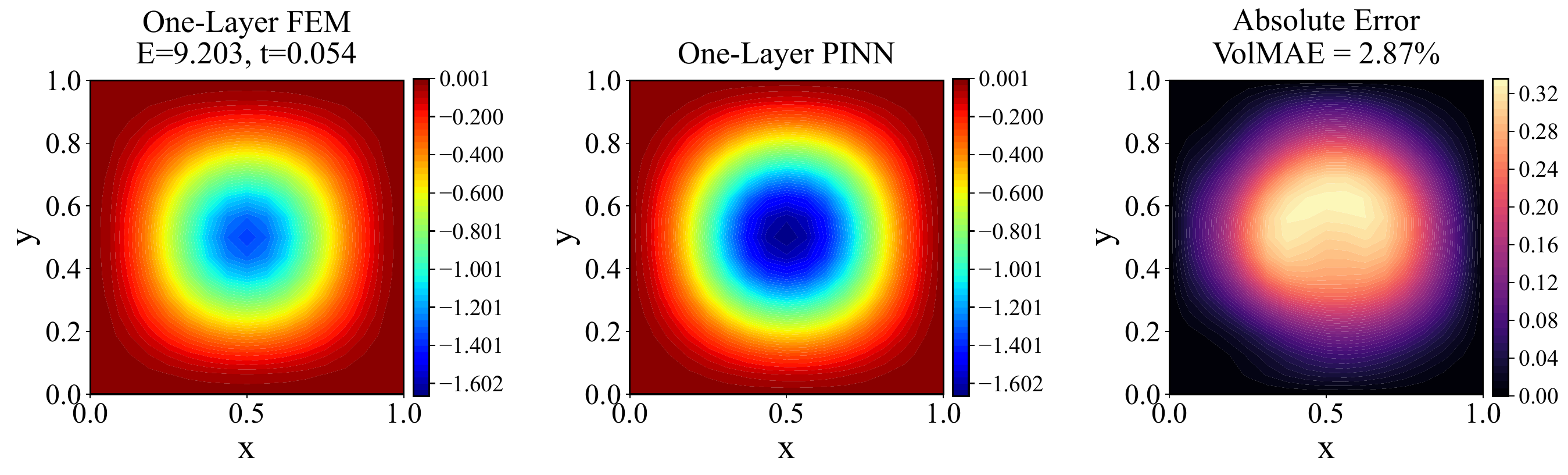}\\[6pt]
\includegraphics[width=\columnwidth]{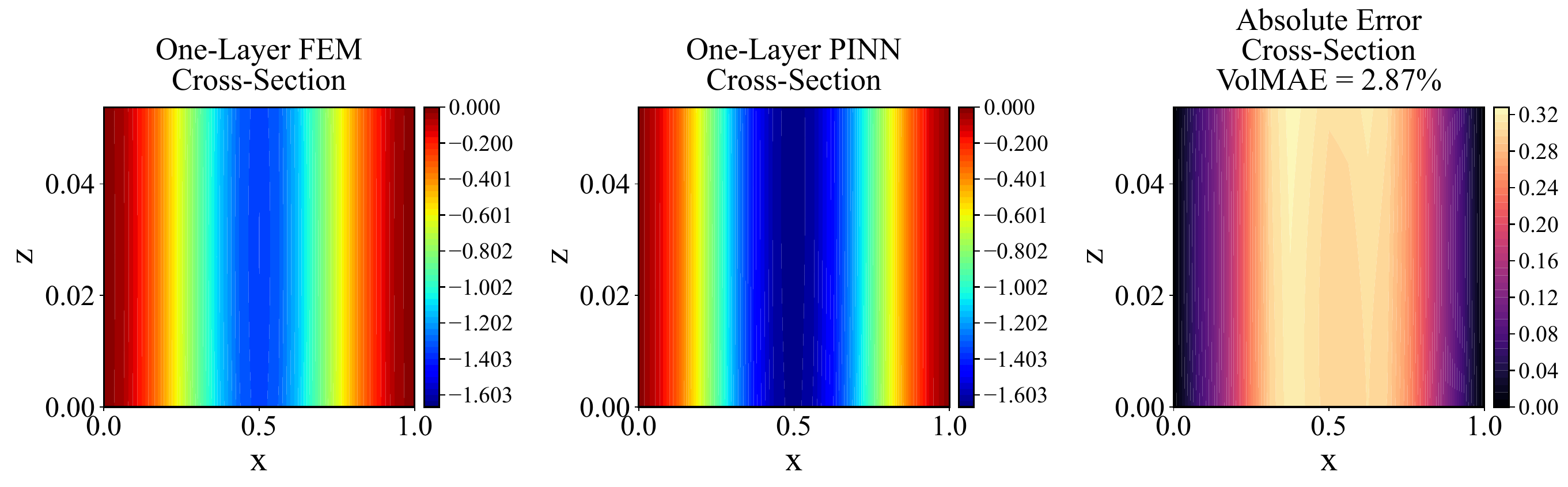}
\caption{Representative PINN-vs-FEM displacement comparisons for single-layer cases. The top row shows the top-surface displacement contour for a representative configuration, followed by the cross-sectional views highlighting agreement between the PINN and FEM solutions through the structure.}
\label{fig:1_PINN_fem_compare}

\vspace{10pt}

\includegraphics[width=\columnwidth]{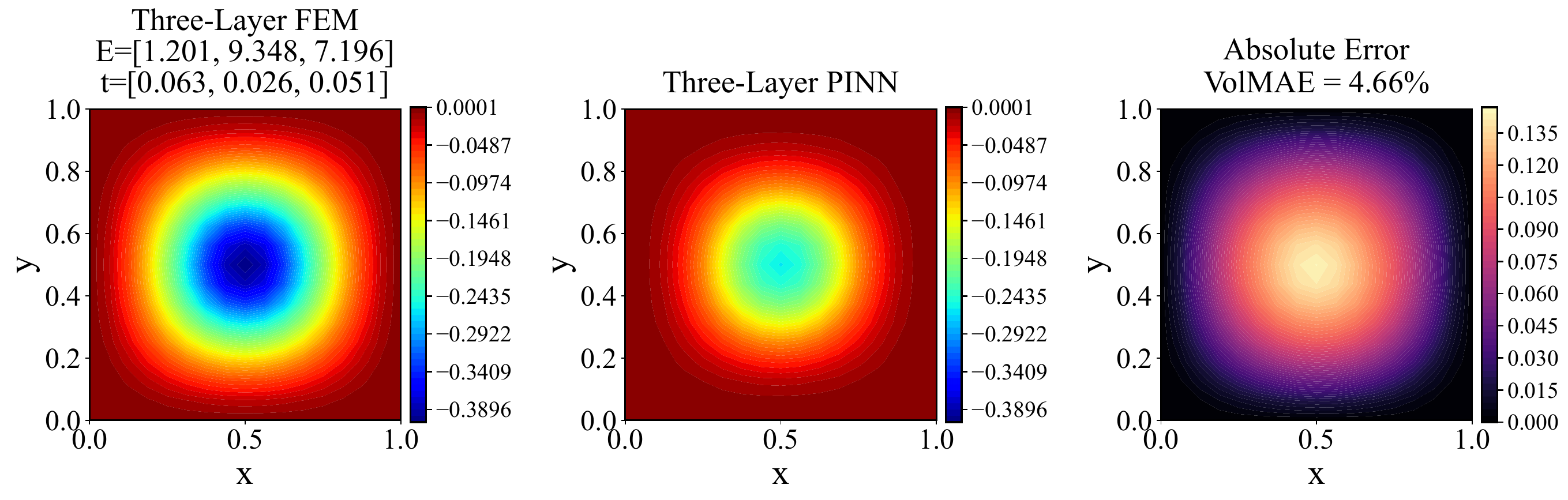}\\[6pt]
\includegraphics[width=\columnwidth]{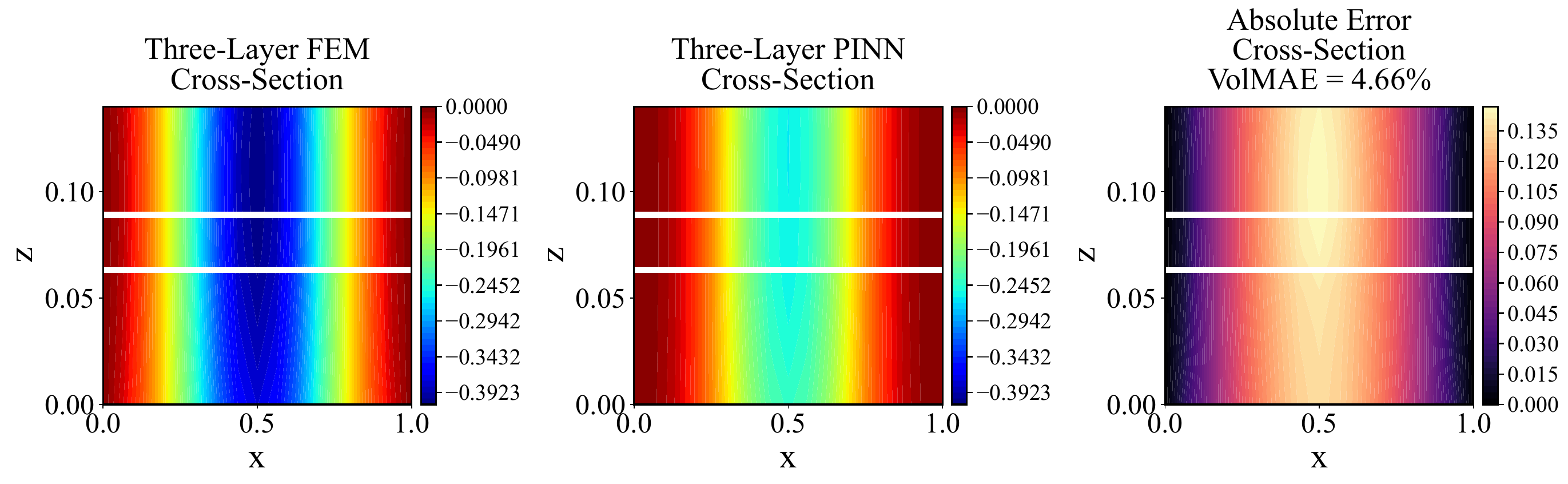}
\caption{Corresponding figure for the three-layered model.}
\label{fig:3_PINN_fem_compare}

\end{figure}

These results indicate that the framework remains accurate even under thin-stack and high-stiffness combinations that are typically difficult for parametric PINNs, proving the effectiveness our method of a unified workflow with defined loss terms provides. Figure~\ref{fig:1_PINN_fem_compare} compares the one-layered models with the FEM while Figure~\ref{fig:3_PINN_fem_compare} visually compares three-layered model with the FEM solution.

\subsection{Ablation Study}

To isolate the effect of the major framework components, we conducted an ablation study to evaluate the  contribution of compliance-aware scaling, layerwise residual decomposition, interface continuity enforcement, and sparse FEM supervision. The results, measured using the same volume-MAE definition as the primary verification studies, for single-layer and three-layer models are presented as followed.

Table~\ref{tab:ablation_three_layer} shows which components are most responsible for the observed improvement in worst-case verification error for the three-layer configuration. The three-layer ablation uses the same documented configuration file as the primary verification study, with 100 random interior parameter configurations, a $16\times16\times8$ FEM mesh, and the same volume-MAE definition. Every ablation variant is trained from the same initialization, with one component removed independently in each row. Removing compliance-aware scaling, sparse FEM supervision, or all proposed components substantially increases volume MAE. Removing layerwise PDE decomposition produces a slightly lower error, while removing interface continuity produces a comparable result. These small differences are interpreted as training variability rather than monotonic improvements from each component.

\begin{table}[htbp]
\caption{Ablation study of the three-layer model.}
\label{tab:ablation_three_layer}
\centering
\begin{tabular}{@{}lcc@{}}
\toprule
\textbf{Variant} & \textbf{Mean MAE (\%)} & \textbf{Worst MAE (\%)} \\
\midrule
\textbf{Full framework} & \textbf{5.78} & \textbf{7.35} \\
\midrule
$-$ Compliance-aware scaling & 31.73 & 119.12 \\
$-$ Layerwise PDE decomposition & 5.39 & 7.03 \\
$-$ Interface continuity & 5.84 & 7.38 \\
$-$ Sparse FEM supervision & 7.89 & 9.02 \\
Vanilla PINN baseline & 30.23 & 129.93 \\
\bottomrule
\end{tabular}
\end{table}

All three-layer ablation variants were independently trained from the same initialization using a fixed random seed and evaluated on the same 100 random interior configurations. The full-framework ablation row is therefore a controlled retraining rather than the selected final checkpoint reported in Table~\ref{tab:PINN_results}. The difference between these results reflects training-run variability and checkpoint selection.

Table~\ref{tab:ablation_one_layer} also displays the components most responsible for the observed improvement in the one-layer configuration. This table uses the same documented configuration file as the primary one-layer verification study, with 100 random interior configurations, a $16\times16\times8$ FEM mesh, all three displacement components, and the same parameter cases and volume-MAE definition. Again, each row is independent of the others, so each row only removes its corresponding component from the full framework. Since the one-layer model is physics-driven and does not need to account for logic between layers, layerwise methods and controlled supervision have been removed.

\begin{table}[htbp]
\caption{Volume-MAE ablation study of the one-layer model.}
\label{tab:ablation_one_layer}
\centering
\begin{tabular}{@{}lcc@{}}
\toprule
\textbf{Variant} & \textbf{Mean MAE (\%)} & \textbf{Worst MAE (\%)} \\
\midrule
\textbf{Full framework} & \textbf{1.56} & \textbf{2.87} \\
\midrule
$-$ Compliance-aware scaling & 2.89 & 6.14 \\
Vanilla PINN baseline & 8.31 & 9.62 \\
\bottomrule
\end{tabular}
\end{table}

\subsection{Visualization and Qualitative Assessment}

Representative qualitative outputs include top-surface displacement contours, cross-sectional displacement profiles, and error heatmaps. Cross-sectional plots are particularly useful for visualizing interface continuity, while error maps reveal where the model is least accurate. These qualitative views complement the scalar MAE metrics by identifying the spatial structure of error and by showing whether the learned field remains mechanically plausible across layer transitions. The following are heat maps of the extremities of the models, both in terms of its top-surface absolute error and cross-sectional absolute error across a mean for randomly-selected parameter values.

\begin{figure}[htbp]
\centering
\includegraphics[width=0.95\linewidth]{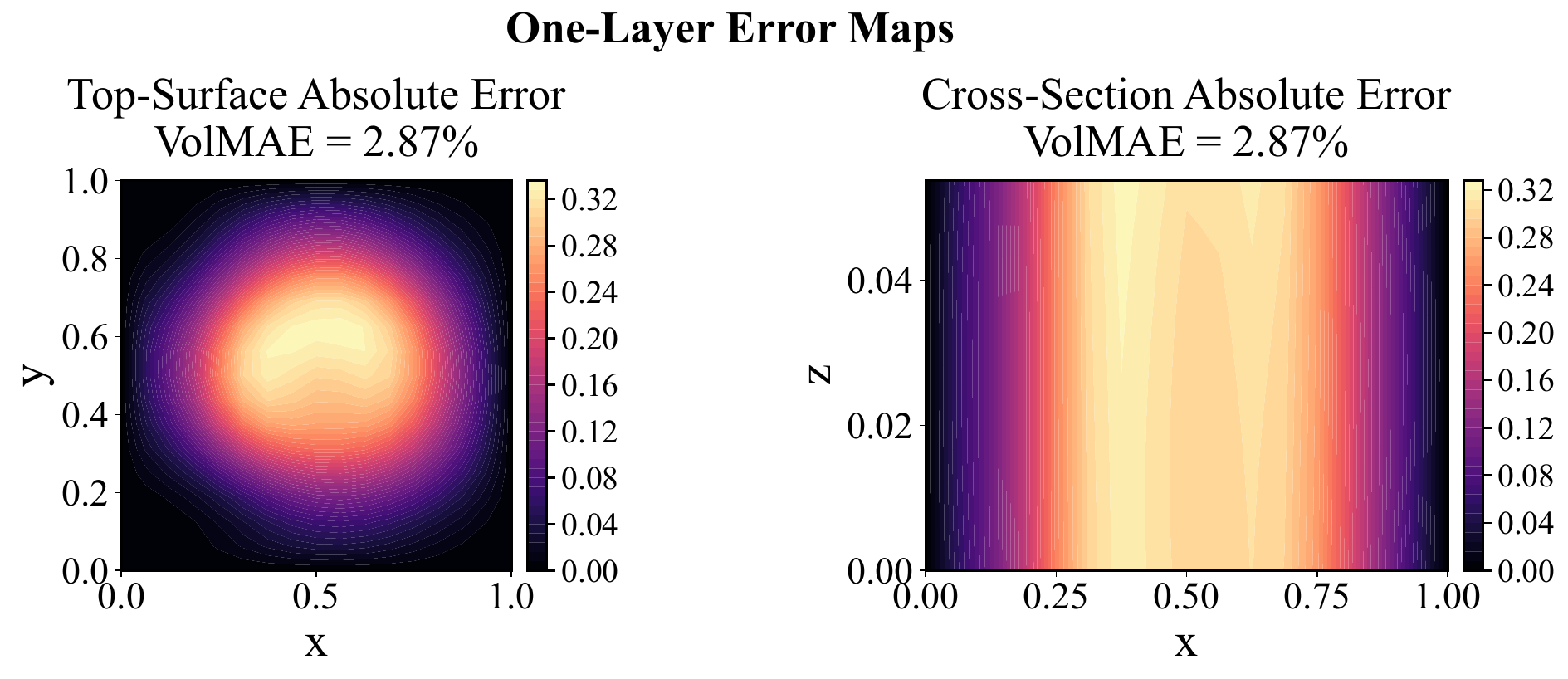}
\caption{Spatial error structure for a representative one-layer verification case. Left: top-surface absolute error $|u_z^{\mathrm{PINN}}-u_z^{\mathrm{FEM}}|$. Right: cross-sectional absolute error through the thickness.}
\label{fig:error_heatmap_one}
\end{figure}

\begin{figure}[htbp]
\centering
\includegraphics[width=0.95\linewidth]{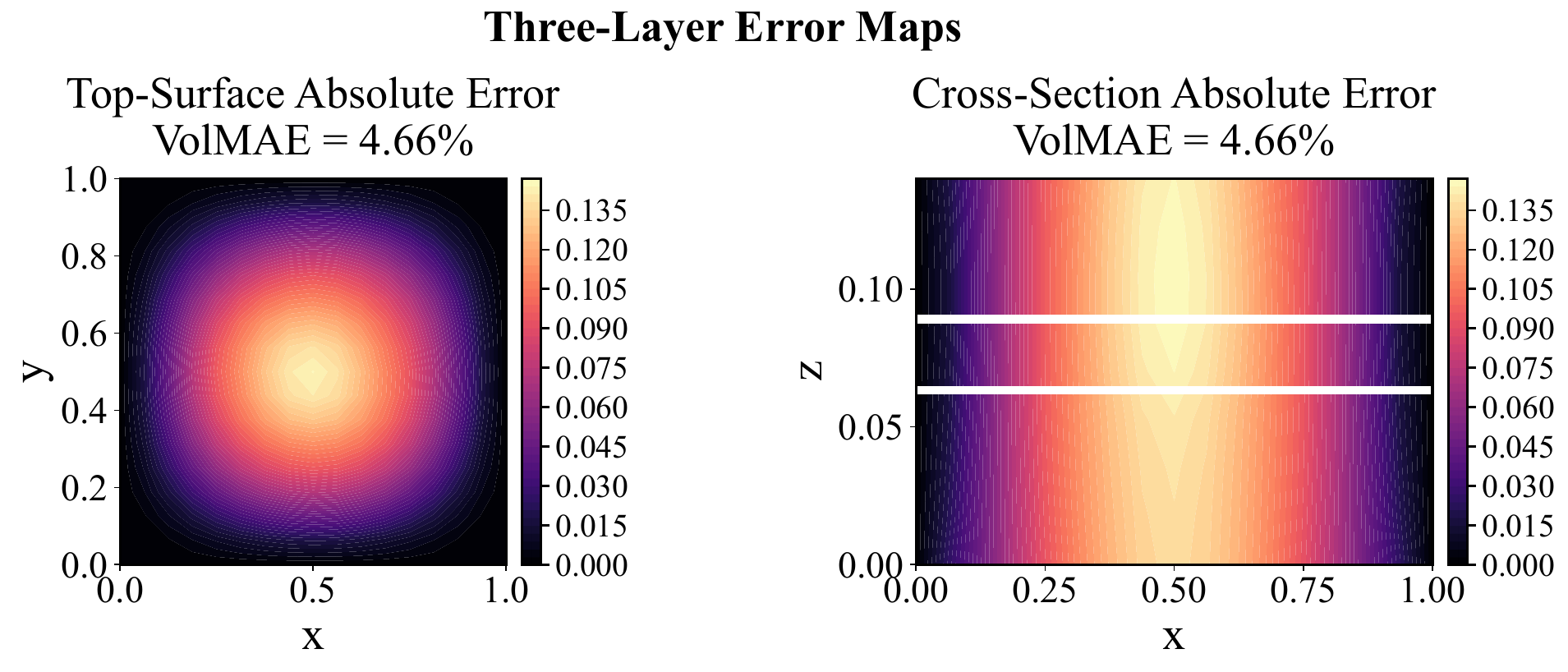}
\caption{Corresponding figure for a representative three-layer verification case. Bold white lines indicate layer interfaces}
\label{fig:error_heatmap_three}
\end{figure}

\begin{figure}[htbp]
\centering
\includegraphics[width=0.95\linewidth]{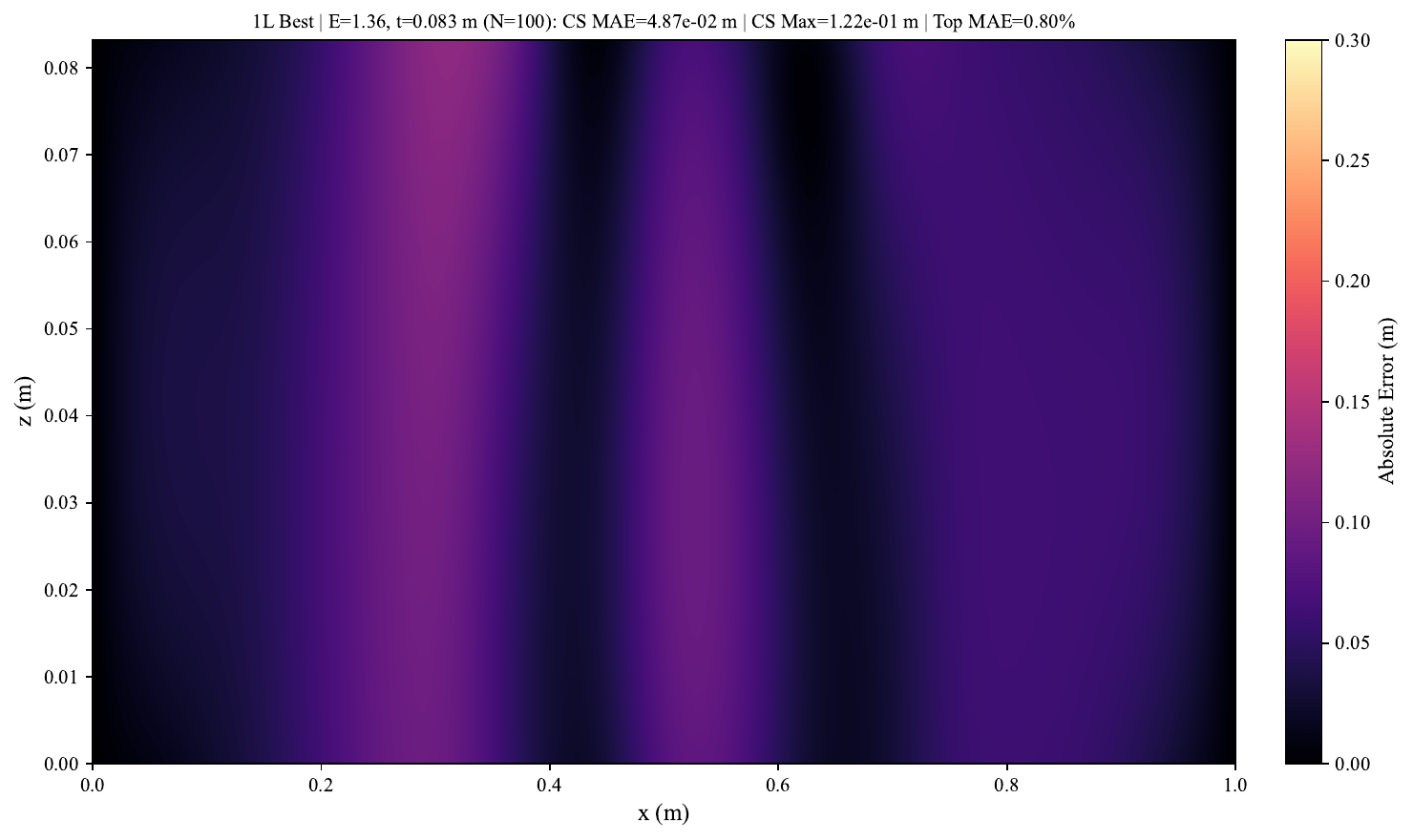}
\includegraphics[width=0.95\linewidth]{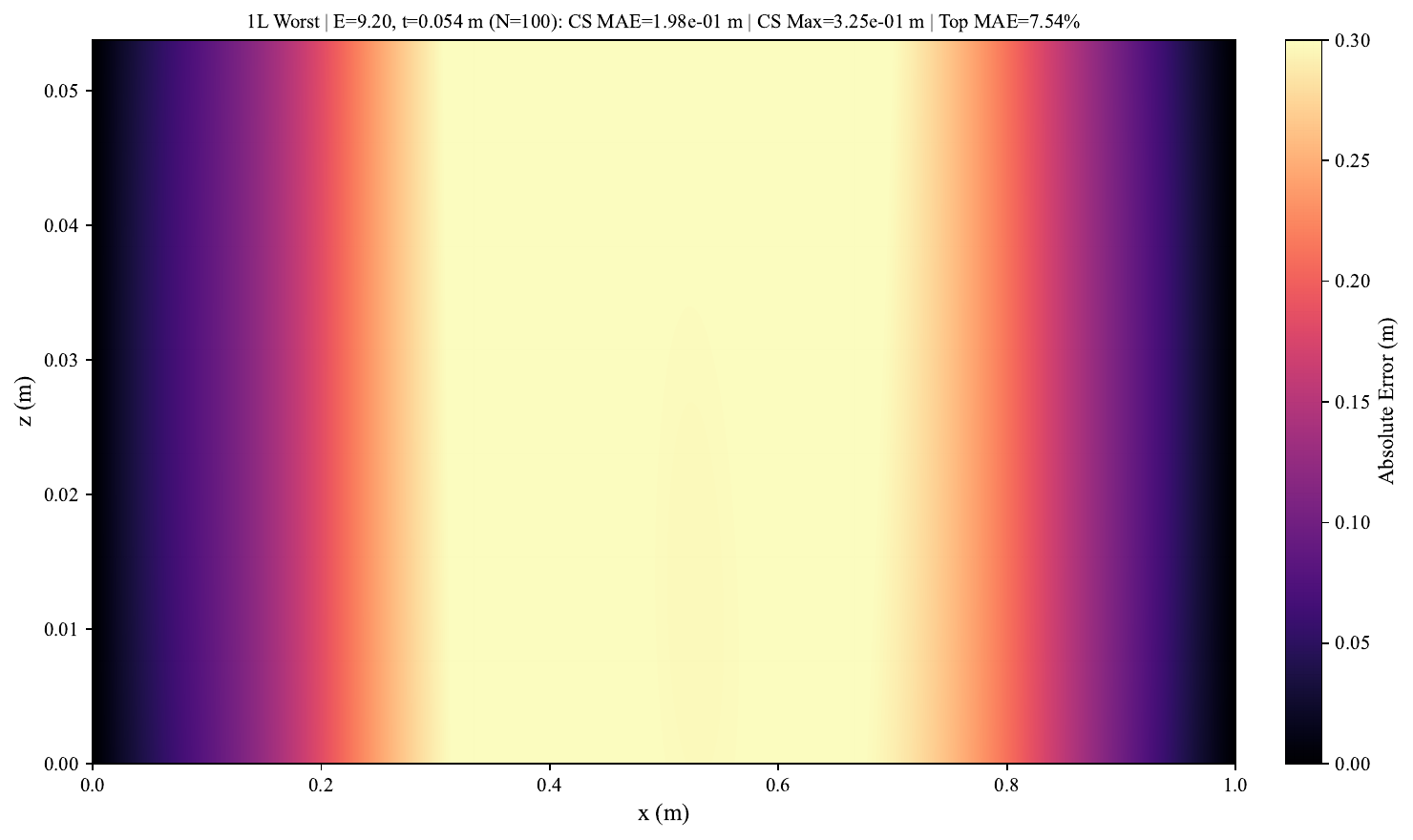}
\caption{Best vs. Worst cross-section error is calculated with $\text{MAE} = \frac{1}{N_{xz}} \sum_{i,k} \varepsilon(x_i, z_k)$}
\label{fig:internal_heatmap_one}
\end{figure}

\begin{figure}[htbp]
\centering
\includegraphics[width=0.95\linewidth]{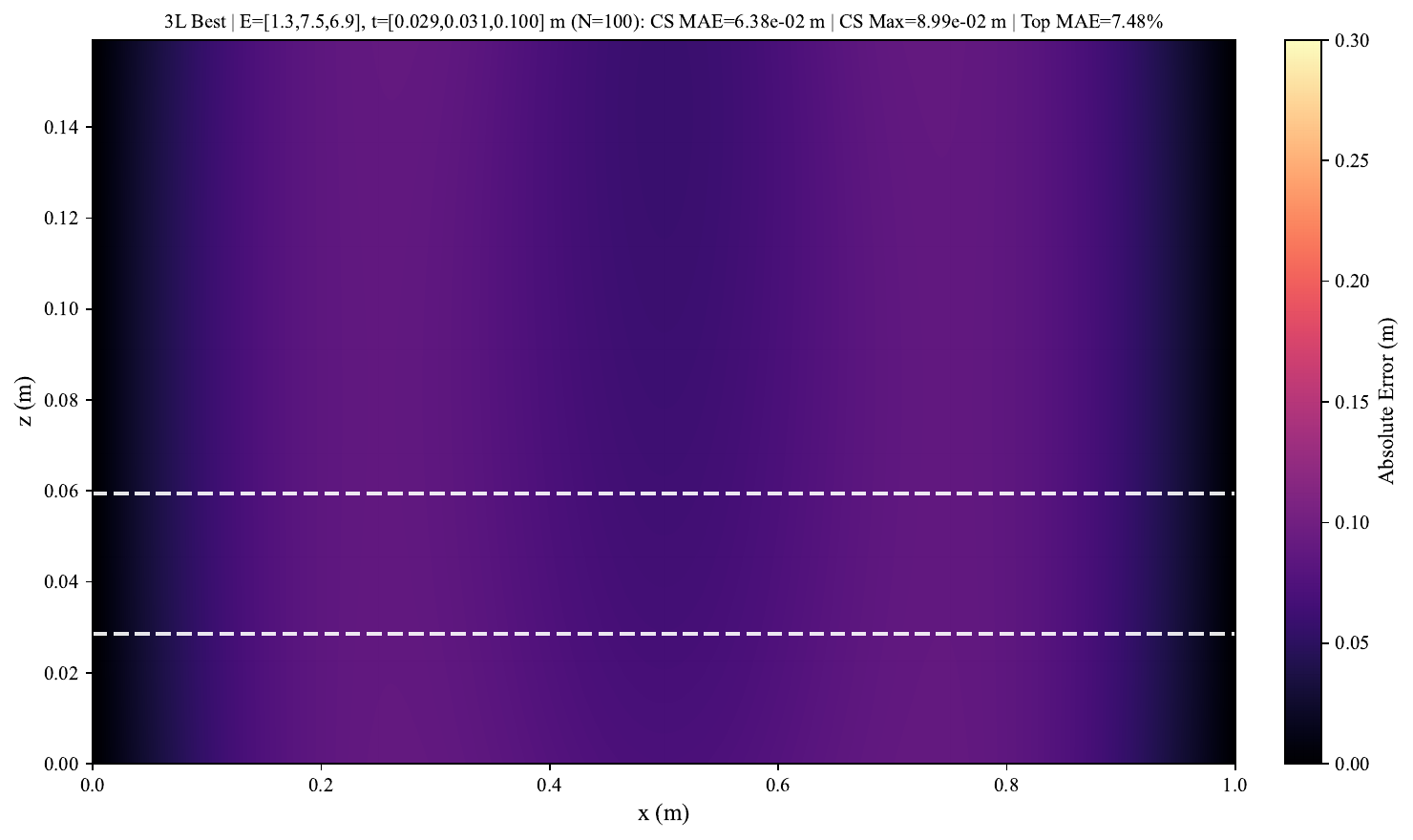}
\includegraphics[width=0.95\linewidth]{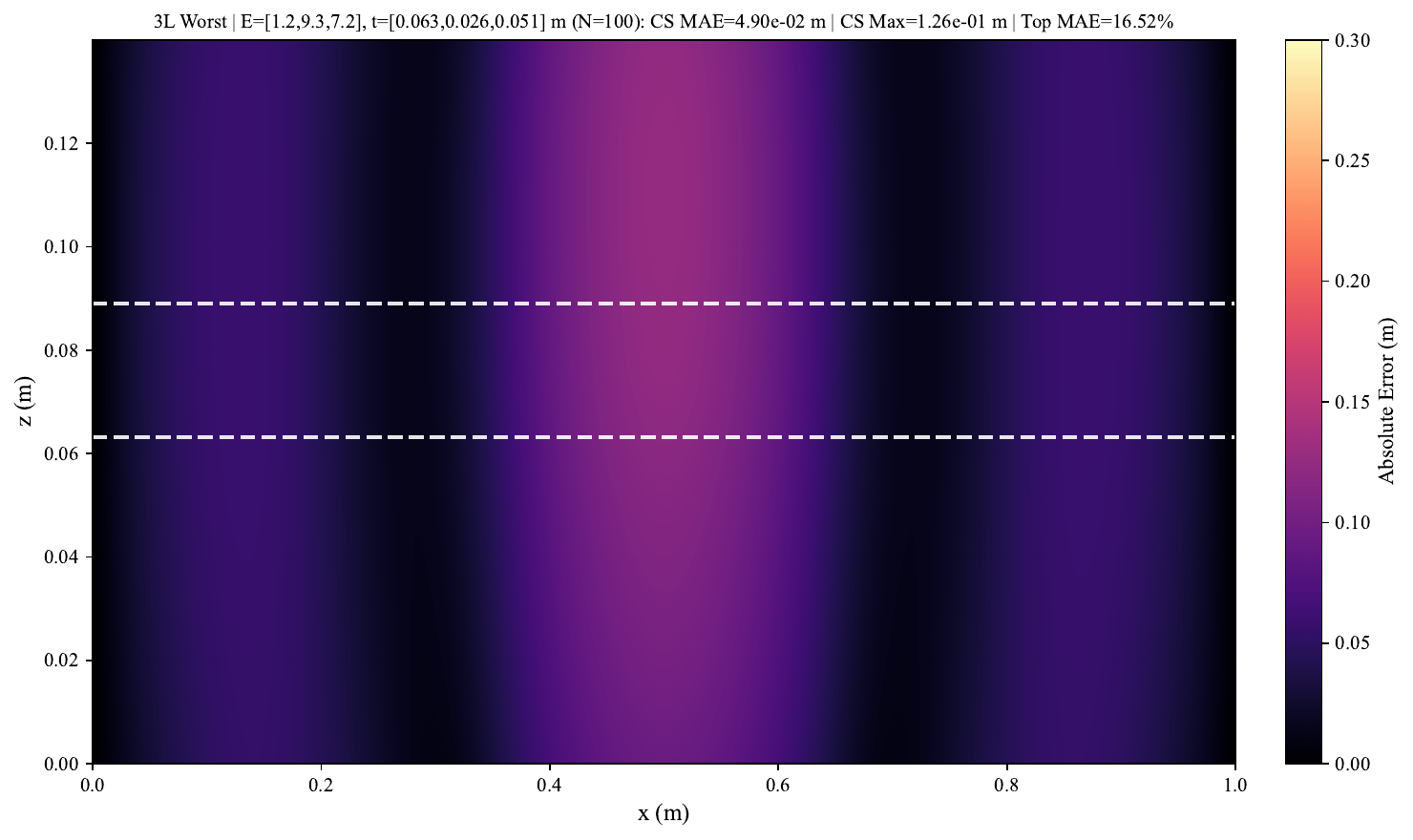}
\caption{Best vs. Worst cross-section error for the three-layer model.}
\label{fig:internal_heatmap_three}
\end{figure}

It is also important to view a generalized study not just the extremities of each interface, but also the internal points where the internal logic is formulated from a pure-physics driven method. Figure~\ref{fig:internal_heatmap_one} and \ref{fig:internal_heatmap_three} present a best-case and worst-case error heatmap of absolute MAE for randomly-chosen configurations of the one-layer and three-layer models. The error patterns are proven to be consistent and representative of the model's behavior across the full parameter space and the interiors of the physical design space. Additionally, the following are two tables (\ref{tab:random_interior}) filled with the top five random interior generalization cases for both the one-layer and the three-layer model.

\begin{table}[!t]
\caption{Five Results of Random Interior Generalization Cases}
\label{tab:random_interior}
\centering
\footnotesize
\setlength{\tabcolsep}{3pt}
\subfloat[One-Layer Model]{%
\begin{tabular}{cccc}
\toprule
Trial & $E$ & $t$ & Vol MAE (\%) \\
\midrule
I   & 6.12 & .135 & 1.67 \\
II  & 3.55 & .133 & 1.49 \\
III & 1.04 & .054 & 2.13 \\
IV  & 5.76 & .113 & 1.32 \\
V   & 4.71 & .143 & 1.72 \\
\bottomrule
\end{tabular}%
}
\vspace{1em}
\subfloat[Three-Layer Model]{%
\begin{tabular}{cccc}
\toprule
Trial & $\mathbf{E}$ & $\mathbf{t}$ & Vol MAE (\%) \\
\midrule
I   & $[6.12, 8.68, 3.55]$ & $[.087, .020, .023]$ & 2.17 \\
II  & $[5.76, 6.65, 4.71]$ & $[.094, .069, .051]$ & 2.33 \\
III & $[7.39, 3.56, 7.95]$ & $[.036, .036, .049]$ & 2.15 \\
IV  & $[2.28, 5.84, 4.67]$ & $[.074, .100, .032]$ & 2.84 \\
V   & $[4.62, 1.02, 8.84]$ & $[.074, .050, .073]$ & 3.50 \\
\bottomrule
\end{tabular}%
}
\end{table}

Both the absolute volume and absolute error reveal that the model internally learns physics relatively accurately. For the one-layered configurations, the absolute cross-section error remains fairly accurate. In the best case scenario, the PINN's interior prediction error remains at 0.87\%, while its worst counterpart has larger discrepancies, with predictions differing by 0.33 and reaching a prediction error of 2.87\%. In the instance of the three-layered configurations, the model and the FEM prediction differs under 0.15, which reveals that for the best case scenario, the error remains at 2.09\%, a satisfactory result, while being 4.66\% for the worst case. In other words, as top layers become thinner, the model's interior predictions deteriorates. This issue most likely arose because of a lack of longer-term training and could be further resolved in future developments.

\section{Conclusion}
\label{sec:conclusion}

We have presented an end-to-end workflow for parameterized physics-informed neural networks applied to a weighted multilayer elastic plate mechanics problem. The framework integrates a trusted FEM baseline, a parametric PINN field model with physics-driven training, automated FEM-referenced evaluation, and a rapid forward call that ensures computational efficiency.

For one-layer configurations, the PINN achieves a mean volume MAE of 1.56\% and a worst-case volume MAE of 2.87\% against FEM. The worst-case configuration occurs at $E=9.203$, $t=0.05375$. For three-layer configurations, through systematic physics-first tuning including layerwise PDE decomposition, strengthened interface sampling, compliance-aware scaling, and lightweight cached supervision, we achieved a final configuration with 2.53\% mean volume MAE and 4.66\% worst-case volume MAE, meeting the stated near-5\% worst-case volume-MAE target for preliminary design space exploration. The worst-case configuration occurs at $E=[1.201, 9.348, 7.196]$, $t=[0.06322, 0.02582, 0.05095]$.

We believe the workflow is technically meaningful and reproducible for multilayer design-space explorations in the given design space with the configuration range. Broader robustness claims should remain tied to the tested sweep and extreme-case settings. The \href{https://github.com/hanb0i/Optimizing-Parameterized-PINNs-to-Efficiently-Solve-Multilayered-Static-Linear-Elastic-PDEs}{GitHub Repository} is publicly available, enabling independent verification and extension.

\section{Future Works}
\label{sec:future_works}

Even though our model has successfully reached the near-5\% worst-case volume-MAE benchmark, it does not suffice for the majority of real-world simulations which encompass sudden spikes of energy from collisions or nonlinear, time-dependent systems. Thus, one future area of development could be geared towards incorporating nonlinearity into the model's training process. Extensions to finite-strain hyperelasticity could leverage energy-based PINN formulations \cite{klein2022hyperelastic} or Deep Ritz variants \cite{weinan2018deep} for large-deformation equilibrium. The current static formulation cannot capture transient stress waves or evolving contact interfaces that arise during actual collisions. Auto-regressive physics-constrained networks have shown promise for dynamic PDE systems \cite{geneva2020modeling}. However, mixed-variable PINN formulations have demonstrated success for elastodynamic wave propagation without labeled data \cite{rao2021elastodynamics}, and iterative $\delta$-PINN refinement has shown promise for correcting transient solution errors \cite{costabal2024deltaPINNs}.

Currently, for higher-dimensional models such as the three-layer PINN, controlled supervised learning from FEM labels is necessary for solution accuracy. Thus, future areas could include curriculum learning and transfer learning strategies to progressively reduce dependence on FEM supervision. Curriculum transfer learning enables convergence without increasing network depth \cite{lyu2024transfer}. Meta-learning approaches that allocate collocation points based on difficulty \cite{toloubidokhti2023dynamic} allow the PINN to extend from single-layer baselines to multilayer systems with minimal FEM data. Meta-learning PINN loss functions provides another avenue \cite{psaros2022uncertainty}. Multifidelity frameworks improve training efficiency \cite{penwarden2023multifidelity}.

While the current FEM-referenced evaluation provides empirical confidence bounds across the tested parameter sweep, uncertainty quantification is not explicit. Therefore, some other areas for improvement include incorporating uncertainty quantification and Bayesian approaches. Bayesian PINNs \cite{yang2021b} could be leveraged to propagate uncertainty through network weights, quantifying prediction confidence. Quantifying total uncertainty \cite{zhang2019quantifying} can address both aleatoric and epistemic sources. Meta-learning approaches \cite{psaros2022uncertainty} could adapt loss functions per problem characteristic, while Bayesian layers or ensemble methods improve uncertainty estimation.

Lastly, to compare our model to accurate FEM, future implementations could incorporate mesh-based coupling. Variational PINNs enforce the weak form via test functions and quadrature, integrating directly with FEM basis functions \cite{kharazmi2021hpvpinn}. Hybrid PINN-FEM methods decompose the domain into FEM regions near boundaries and PINN regions in the interior, outperforming pure PINNs on complex geometries \cite{sobh2025pinnfem}. Wave propagation has been addressed with physics-informed formulations \cite{rasht2021wave}. Examples such as $\delta$-PINNs \cite{costabal2024deltaPINNs} have achieved fine-mesh FEM accuracy at reduced cost.

\bibliographystyle{IEEEtran}
\bibliography{ref}

\end{document}